# Condensation of electron-hole pairs in a degenerate semiconductor at room temperature

Peter P. Vasil'ev [*]

*PN Lebedev Physical Institute, 53 Leninsky Prospect, Moscow 119991, Russia,*

also with: *Central Research Laboratory, Hamamatsu Photonics K.K., 5000 Hirakuchi, Hamamatsu City, 434 Japan*

Igor V. Smetanin

*PN Lebedev Physical Institute, 53 Leninsky Prospect, Moscow 119991, Russia*

Abstract

It has been theoretically shown that in large-density semiconductor plasma there exist an energy level of a bound electron-hole pair (a composite boson) at the band gap. Filling this level up occurs through the condensation of electron-hole pairs with the use of mediating photons of a resonant electromagnetic field. We have demonstrated that in the case of a strong degeneracy of the plasma the critical temperature of the condensation is determined by the Fermi energies of the plasma components rather than the order parameter $\Delta$. The critical temperature can exceed 300 K at electron-hole densities as large as $6 \cdot 10^{18}$ cm$^{-3}$. The theoretical model is consistent with available experimental data.

## I. INTRODUCTION

The problem of condensation of electron-hole pairs in semiconductors have recently attracted a great deal of research interest[1,2]. The Coulomb attraction ensures the formation of excitons, which may condense and obey the Bose statistics at low densities and low enough temperatures. In the large density case, collectively paired electrons and holes may form BCS-like state that is quite similar to the ensemble of Cooper pairs in a superconductor. The possibility of condensation in such systems was first discussed by L.V.Keldysh with co-workers[3,4]. The *e-h* condensation has been widely studied theoretically [1,2,5-14]. Bose condensation of excitons is believed to be experimentally observed in some semiconductors and quantum-well structures [15-17].

It is generally accepted that electron-hole plasma can be created in a semiconductor by external optical pumping. The behaviour of electrons, holes, and quasiparticles in strong resonant electromagnetic fields was first considered by Galitzkii *et al* [18,19]. A model of a semiconductor driven by a strong classical monochromatic laser field was investigated by Comte and Mahler[7]. It was demonstrated that the presence of the laser field resulted in a large enhancement of the order parameter. Spectra of absorption and luminescence of electron-hole condensate were theoretically studied in many papers [13, 20-22]. Lasing effect from coherent excitonic states in quantum-well structures was discussed by Flatte *et al* [23].

Superradiant emission in large-density nonequilibrium *e-h* systems in GaAs/AlGaAs *p-i-n* heterostructures was experimentally studied at room temperature [24-28]. The carrier density exceeded $(2-6) \cdot 10^{18}$ cm$^{-3}$ in the experiments. Large intensity (~ $10^9$ W/cm$^{-2}$) femtosecond superradiant emission pulses were generated. The superradiant pulses were emitted from a narrow spectral bandwidth corresponding to the energy of the renormalized band gap[24,25]. It was shown that extremely large peak powers (energies) generated from a narrow spectral range at femtosecond pulsewidths could not be explained by the standard Fermi-Dirac distribution of electrons and holes in the bands and the square root dependence of density of states [27]. It was directly observed [28] how electrons and holes moved from energy levels inside the bands towards the bottom of the conduction band and top of the valence band. The condensation of the carriers at the energy levels right at the band gap resulted in the formation of the nonequilibrium BCS-like coherent state, which finally recombined in the form of a femtosecond high-power superradiant pulse. It is worth to point out here that it is not possible to explain the results of the experiments[24-28] in terms of the conventional theory of semiconductor lasers.

The main difficulty of the interpretation of available experimental data consists in the fact that the estimated order parameter $\Delta \sim$ 1-2 meV is much smaller compared with the sample temperature T = 25 meV. One of the goals of the present paper is to overcome this problem. Here, we demonstrate that the critical temperature of the condensation at the lowest energy level is determined by the Fermi energy in case of a highly degenerate *e-h* system rather than the order parameter in case of superconductivity. Electron-hole pairs condense at the bottoms of the bands. When a pair is broken up, the electron and hole must make a transition to a free energy level that exists at the Fermi energy $\mu_{e,h}$ (see Fig. 1). On the other hand, Cooper pairs in a superconductor occupy a region near the Fermi surface. It is enough to have a small energy about the order parameter for breaking them up. In Section 2 of this paper we demonstrate the existence of a low-energy state near the band gap of a bound *e-h* pair (a composite boson) in the large-density case. Section 3 proceeds with a qualitative description of the process of felling up of this level with composite bosons with the assistance of photons of a resonant electromagnetic field. Section 4 gives a theory of the condensation of e-h pairs within the framework of the mean-field approximation. An approximate wave function of the condensate will be given, probabilities for the destruction of pairs and the critical temperature will be calculated.

II. LOW-ENERGY STATE OF A BOUND E-H PAIR IN A SEMICONDUCTOR

The problem of the existence of a low-energy state at the band gap of a bound e-h pair has a great importance for the consideration of the *e-h* condensation in a semiconductor in large-density case at room temperature. Excitons states close to the Fermi energy, arising from the electron-hole Coulomb attraction, in a degenerate semiconductor was studied in Ref. 29. Let us now discuss the existence of a low-energy state.

In this paper we restrict ourselves by a qualitative consideration of this problem and use a simplified model, in which an electron and hole attract by the Coulomb force and moves in a mean field of another particles and lattice. We neglect here any spin states since they are not important [30,31]. This problem is pretty similar to problems of a weakly bound exciton and bound states of Cooper pairs (see, for instance, Ref. 30 and 31, respectively). The Hamiltonian of the electron-hole system looks like

$$H = H_{0e} + H_{0h} - \frac{e^2}{\varepsilon_s |r_e - r_h|}, \tag{1}$$

where

$$H_{0e,h} = -\frac{\hbar^2}{2m_{e,h}} \nabla^2_{re,h} + U_{e,h}(r_{e,h}). \tag{2}$$

Here $U_{e,h}(r_{e,h})$ is the effective potential of interaction between the electron (e) and hole (h) with the lattice and another particles, $r_{e,h}$ is the corresponding radius vector, $\varepsilon_s$ is the dielectric constant of the semiconductor.

The wave function of the *e-h* pair is determined by the solution of the Schrodinger equation with the Hamiltonian (1), (2). If one neglects the interaction of the pair with another electrons and holes, then the problem leads to the classic excitonic solutions of the well-known form with the hydrogen-like atom energy spectrum [30]. By contrast, in the case of the large density case the Coulomb attraction energy of the pair is a small parameter. In accordance with the standard perturbation theory[31] we find the wave function in the form decomposition on the wave functions of free electrons and holes [32]

$$\Psi(r_e, r_h) = \frac{1}{V} \sum_{k,k'} C_{k,k'} \exp[ikr_e + ik'r_h], \tag{3}$$

where the summing up is done over all possible states within the Fermi surface, $V$ is the volume of the crystal. Substituting (3) into the Schrodinger equation, we obtain the dispersion equation for the energy of the pair $E$ as

$$\frac{1}{V} \sum_k \frac{1}{\varepsilon_e(k) + \varepsilon_h(k) - E} = \frac{\varepsilon_s k_{TF}^2}{4\pi e^2}. \tag{4}$$

Here $\varepsilon_{e,h}(k)$ is the eigenvalues of the energy of electrons and holes having the wavevector $k$. We assume that the Coulomb interaction is screened with the Thomas-Fermi wave number $k_{TF}$, and an electron and hole have wavevectors $k$ and $-k$.

The dispersion equation (4) has a solution $E_0$, which lies below the band gap. This solution corresponds to the bound state of the electron and hole. Indeed, in the range $\varepsilon_e(k_{max}) + \varepsilon_h(k_{max}) > E > \varepsilon_e(k_{min}) + \varepsilon_h(k_{min})$, with $k_{max}$ and $k_{min}$ are maximum and minimum allowed wavevectors, the left part of Eq. (4) oscillates quickly and has multiple resonances. By contrast, in the range $E < \varepsilon_e(k_{min}) + \varepsilon_h(k_{min})$ it is positively defined and asymptotically decreases when $E \to -\infty$. This implies that the Eq. (4) do have a solution in the required range.

In terms of a simplified model of a semiconductor with the isotropic parabolic bands one has

$$\varepsilon_e(k) + \varepsilon_h(k) = \frac{\hbar^2 k^2}{2m_e} + \frac{\hbar^2 k^2}{2m_h} + E_g, \tag{5}$$

where $E_g$ is the band gap in an infinite crystal. A ratio of heavy and light holes in intrinsic GaAs, used in the experiments Ref. 24-28, is around 7:1, the heavy hole mass being $m_h \approx 0.5 m_0$, the effective electron mass $m_e \approx 0.067 m_0$, with $m_0$ being mass of the electron. We neglect the movement of the holes and do summing in Eq. (4) by electron state in the conduction band. Replacing the sum by the integration, we have the following form of the dispersion equation

$$\frac{\sqrt{2}}{\pi^2}\frac{m_e^{3/2}}{\hbar^3}\int_{\chi_{min}}^{\chi_{max}}\frac{\sqrt{\chi}d\chi}{\chi+E_g-E_0}=\frac{\varepsilon_s k_{TF}^2}{4\pi e^2},\qquad(6)$$

where $\chi=\hbar^2 k^2/2m_e$. The lower integration limit corresponds to the minimum possible wave number $k_{min}$, which is determined by the dimensions of the crystal. The upper limit is chosen to be determined by the Thomas-Fermi wave number $k_{TF}=[16 m_e e^2/(\pi\varepsilon_s)](6\pi^2 n)^{-1/3}$.

Solutions of Eq. (6) differ strongly depending on the sign of $E_g-E_0$. When the bound *e-h* state lies below $E_g$ for an infinite crystal, we have after the integration an implicit algebraic equation

$$1-\delta-\sqrt{x}\left[\arctan\left(1/\sqrt{x}\right)-\arctan\sqrt{\delta/x}\right]=\frac{(18\pi)^{1/3}}{r_s}\left(Ry/\chi_{max}\right)^{1/2},\qquad(7)$$

with $x=(E_g-E_0)/\chi_{max}$, $\delta=\chi_{min}/\chi_{max}\ll 1$, $Ry$ is the exciton Rydberg and $r_s$ is the mean interparticle distance in the units of the exciton Bohr radius.

It is clear that the function on the left hand side of Eq. (7) has values in the range of (0,1). When the *e-h* density increases, i.e. $r_s$ decreases, the value of the right hand side of Eq. (7) increases, Eq. (7) being unsolvable starting from a certain value $r_{scr}$. In order to estimate $r_{scr}$ let us assume that $\chi_{min}\sim 0$. Then we have for the energy of the bound state $E_0$

$$\frac{E_g-E_0}{\chi_{max}}\approx\left(4/\pi^2\right)\left[1-\frac{(3\pi^2)^{1/3}}{2\sqrt{2}}r_s^{-1/2}\right]^2.\qquad(8)$$

Using (8) one has

$$r_{scr}\approx 8/(3\pi^2)^{2/3}\approx 0{,}84\qquad(9)$$

At $r_s = r_{scr}$ $E_0$ is equal to $E_g$. For $r_s < r_{scr}$ the dispersion equation (6) must be solved assuming $E_0 > E_g$. However, values of $\chi$ should not exceed $\chi_{min} = \dfrac{\hbar^2 (\pi/d)^2}{2m_e}$, where $d$ is the minimum dimension of the crystal.

In this case the integration gives the following relation for the dimensionless parameter $y = -x = (E_0 - E_g)/\chi_{max}$

$$1 - \delta + \frac{\sqrt{y}}{2} \ln \left| \frac{(1-\sqrt{y})(\sqrt{\delta}+\sqrt{y})}{(\sqrt{\delta}-\sqrt{y})(1+\sqrt{y})} \right| = \left( \frac{r_{scr}}{r_s} \right)^{1/2}. \quad (10)$$

This equation can be readily studied at $y \to 0$ and $y \to 1$. For instance, for $r_s \leq r_{scr}$ and small $(E_0 - E_g)$ we have

$$\frac{E_0 - E_g}{\chi_{max}} \approx \sqrt{\delta} \left[ \frac{1}{1-\sqrt{\delta}} \left( \frac{r_{scr}}{r_s} \right)^{1/2} - 1 \right]. \quad (11)$$

This relationship is valid for a narrow density range $1 - r_s/r_{scr} \ll 2\sqrt{\delta}$. For example, for the densities under consideration and the crystal thickness of about 0.1 μm $\sqrt{\delta} \sim (k_{TF} d)^{-1} \sim 10^{-2} \div 10^{-3}$. When the *e-h* density increases further, the energy of the bound state asymptotically approached $E_g$

$$\frac{E_0 - E_g}{\chi_{min}} \approx 1 - 2 \exp\left( -\frac{2}{\sqrt{\delta}} \left[ \sqrt{\frac{r_{scr}}{r_s}} - 1 + \delta \right] \right) \quad (12)$$

Note that the solutions (10)-(12) disappear in the case of a large crystal $\chi_{min} \to 0$.

Dependencies of $E_0$ as a function of $r_s$ are illustrated in Fig. 2. Figure 2(a) shows the solution of Eq. (7) which is valid for densities smaller that the critical one $r_s > r_{scr}$. It is worth to mention here that the extrapolation of solutions of Eq. (7) into the range of small densities does not have solutions in the form of excitons. Indeed, the dispersion equation (4) was derived as a result of the decomposition on the wave functions of unbound electrons and holes (3). This assumption is valid at large enough densities, when a perturbation, which is produced by neighbour electrons and holes, substantially affects the spectrum of the bound pair. $E_0$ against $r_s$ at densities larger than critical is plotted in Fig. 2 (b) for different values of the parameter $\delta \sim (k_{TF} d)^{-2}$. These dependencies illustrate a fast asymptotic approach of the bound state to the bottom of the band with increasing density.

The present analysis has shown the existence of bound states in a dense *e-h* ensemble. These states locate at energies just below the band gap $E_g$. Now let us discuss in details the process of filling up of this bosonic state.

III. MECHANISM OF CONDENSATION ONTO BOSONIC LEVELS

Here we consider the following mechanism of the formation of the condensate. It has been proposed earlier in Ref. 27 and 28. As a result of spontaneous radiative recombination of electrons and holes in the active layer of a GaAs/AlGaAs heterostructure, a macroscopic electromagnetic field develops at the narrow spectral range right at the band gap. The corresponding profile of the optical gain [27] can be provided by a proper reverse bias on the absorber of the structure. This ensures the absorption of photons with shorter wavelengths. Consider now an electron and hole that locate at the bottom of the conduction band and top of the valence band, correspondingly. A photon of the electromagnetic field is a resonant one for them (see the initial state in Fig. 3). The photon causes their stimulated radiative recombination and one has two identical photons as an intermediate state in Fig. 3. Due to a very fast intraband thermolization, the energy levels, which became free, are quickly occupied by an electron and hole from levels with higher energy. At the same time as a result of the opposite process, one of the photons generates a bound *e-h* pair, which is coherent for some time with the electromagnetic field. The bound pair occupies the bosonic level discussed in the previous Section (the final state in Fig. 3). The absorption of the photon with the generation of unbound electron and hole does not happen since all the electrons and holes levels at the bottom of the band are occupied. This process takes place again and again.

An ensemble of bound e-h pairs (composite bosons), which have zero total wavevector, develops with the assistance of the resonant electromagnetic field. In addition, this field enforces coherency throughout the forming condensate. Therefore, a photon-assisted drain of particles occurs from fermionic ensemble of unbound electrons and holes to bosonic system of bound pairs. Once again, the process of the condensation is thought to happen due to the emission of photons with recombination of carriers from fermionic states, whereas the absorption of photons occurs on bosonic states of bound pairs. This leads to the condensation in the phase space. Because bound *e-h* pairs are bosons composed with two fermions, they cannot occupy just one energy level. In case of larger density the energy distribution of bound *e-h* pairs resembles that of Cooper pairs (see, for instance Ref. 24,25, and 27).

The break-up of bound pairs occurs due to their scattering by optical phonons with transitions of electrons and hole to free energy levels inside the bands near the Fermi surface[27]. Radiative recombination of a bound pair with the generation of a photon does not result in the destruction of the condensate. This is because the pair is virtually a two-level system, and the photon annihilates with the generation of an identical bound pair within a time interval shorter compared with the round trip time inside the crystal. In a certain time the condensate coherently recombines via the generation of a high-power cooperative (superradiant) femtosecond pulse. The peak power and pulse width is determined by the geometry of the sample and the phase relaxation time $T_2$ [24-28].

It should be noted that without a resonant electromagnetic field the bosonic level is virtually empty because $\Delta \ll kT$. We discuss here the crucial role of the internal resonant electromagnetic field. However, it is clear that this role can be played by any external photon field having a properly chosen energy (frequency) and intensity.

## IV. CRITICAL TEMPERATURE OF THE CONDENSATE

As we mentioned above, the range of densities achieved experimentally[24-28] was $(2 \div 6) \times 10^{18} cm^{-3}$, the corresponding values $r_s \sim 0.31 \div 0.59$. Note that the value of the exciton Rydberg and the dielectric constant in GaAs are $Ry \approx 4$ meV and $\varepsilon_s \approx 13.4$, respectively. For the description of the condensate we use a standard approach based on the Bogolubov transformation, which has been previously developed by different authors[2-14]. However, it is necessary to point out a substantial difference between the physical situation described in those papers and conditions of the experiments in Ref. 24-28. It was supposed in the theoretical papers that electron-hole pairs were created by an external optical field with the photon energy $\hbar\omega_L > E_g$. This fixes the chemical potential of the electron-hole plasma $\mu^*$ due to the relation[7]

$$\mu^* = \hbar\omega_L - E_g \quad . \tag{13}$$

As a result one has a) the resonant interaction of the electromagnetic field occurs with *e-h* pairs that occupy energy level near the Fermi energy; b) the plasma density is determined by the field amplitude and its frequency. These strict connections were absent in the experiments[24-28]. The plasma density and consequently the chemical potential were conditioned by the current injection, whereas the photon energy of the resonant interaction was determined by the profile of the optical gain. The latter was controlled by the voltage bias on the absorber section of devices [27,28]. In this case the condition (13) is not fulfilled and we can have the following conditions

$$\frac{\mu - \hbar\omega}{T} \gg 1, \qquad \frac{\mu - \hbar\omega}{\Delta} \gg 1 \tag{14}$$

right until the very bottoms of the bands.

We restrict ourselves here by consideration of electrons and heavy holes since the number of light holes is relatively small in GaAs. The full Hamiltonian is a sum of the kinetic energy of electrons and holes, the energy of their Coulomb interaction, and the energy of the resonant interaction with the electromagnetic field

$$H_T = H_{kin} + H_{Coul} + H_L . \tag{15}$$

Here

$$H_{kin} = \sum_k E_e(k) a_k^+ a_k + E_h(k) b_k^+ b_k , \tag{16}$$

where $a_k^+, a_k, b_k^+, b_k$ are the operators of the generation and annihilation of electrons and holes with the wave number $k$, correspondingly. Their energies are $E_{e,h}(k) = E_g/2 + \hbar^2 k^2 / 2m_{e,h}$.

The Coulomb part of Eq. (15) is

$$H_{Coul} = \frac{1}{2} \sum_{k,k',q} V_q [a_{k+q}^+ a_{k'-q}^+ a_{k'} a_k + b_{k+q}^+ b_{k'-q}^+ b_{k'} b_k - 2 a_{k+q}^+ b_{k'-q}^+ b_{k'} a_k], \tag{17}$$

where

$$V_q = 4\pi e^2 \hbar^2 / \varepsilon_s q^2 . \tag{18}$$

The last part of Eq. (15) is given by

$$H_L = \sum_k (d_0 E_L)\left(a_k^+ b_{-k}^+ \exp[-i\omega t] + a_k b_{-k} \exp[i\omega t]\right) . \tag{19}$$

Here $d_0$ is the effective dipole matrix element of the interband transition, being weakly dependent on $k$, $E_L$ is the amplitude of the electric field, being monochromatic one with the frequency $\omega$, $E = E_L \sin[k_l r - \omega t]$. Note that we keep in (19) the resonant transitions, the electrons and hole have equal wave numbers $k$ and $-k$ (k >> $k_L$).

The great canonical transformation [18,19] allows us to eliminate the dependence on time in Eq. (16). After the energy renormalization we have the Hamiltonian

$$H = \sum_k (E_e(k) - \hbar\omega/2) a_k^+ a_k + (E_h(k) - \hbar\omega/2) b_k^+ b_k + H_{Coul} + \\ \sum_k (d_0 E_L)(a_k^+ b_{-k}^+ + b_{-k} a_k) \tag{20}$$

We use the Bogolubov transformation[31] for the diagonalisation of the Hamiltonian (20)

$$\alpha_k^+ = u_k a_k^+ + v_k b_{-k}, \qquad \beta_k^+ = u_k b_k^+ - v_k a_{-k} \tag{21}$$
$$u_k^2 + v_k^2 = 1$$

The Fermi operators $\alpha_k^+, \alpha_k, \beta_k^+, \beta_k$ describe new quasiparticles, dressed electrons and holes. After introducing (21) into (20) one gets

$$H = H_0 + H_2 + H_4, \tag{22}$$

where $H_0$ does not contain the operators of the quasiparticles,

$$H_0 = \sum_k [\xi_k v_k^2 - 2(d_0 E_L) u_k v_k] - \sum_{k,k'} V_{k-k'} (u_k v_k u_{k'} v_{k'} + v_k^2 v_{k'}^2), \tag{23}$$

and represents in fact the ground state. $H_2$ looks like

$$H_2 = \sum_k [\varepsilon_{k\alpha} \alpha_k^+ \alpha_k + \varepsilon_{k\beta} \beta_k^+ \beta_k] + \sum_k [\delta_k u_k v_k - (u_k^2 - v_k^2) \Delta_k] (\alpha_k^+ \beta_{-k}^+ + \beta_{-k} \alpha_k), \tag{24}$$

$H_4$ has all possible quadruple multiples of the operators. It describes processes of scattering of quasiparticles. It is not essential for our treatment and will be omitted from now on. Note that the Coulomb terms in $H_4$ correspond to scattering on large angles, whereas the dominant scattering on small angles is taken into account in $H_2$ and $H_0$. In (23), (24) the following parameters are introduced [7]

$$\varepsilon_{k\alpha} = \frac{1}{2}(\varepsilon_k + \varsigma_k), \qquad \varepsilon_{k\beta} = \frac{1}{2}(\varepsilon_k - \varsigma_k) \tag{25}$$

are the quasiparticle energies,

$$\Delta_k = \Delta_k^0 + 2(d_0 E_L), \qquad \Delta_k^0 = 2\sum_{k'} V_{k-k'} u_{k'} v_{k'} \tag{26}$$

is the effective order parameter. The following relations are also valid

$$\xi_k = E_e(k) + E_h(k) - \hbar\omega, \qquad \varsigma_k = E_e(k) - E_h(k),$$
$$\delta_k = \xi_k - 2\sum_{k'} V_{k-k'} v_{k'}^2, \qquad \varepsilon_k = \sqrt{\delta_k^2 + \Delta_k^2}. \tag{27}$$

Here $\xi_k$ is the detuning of the pair with wave numbers $k, -k$ from the resonance, $\varsigma_k$ is characterized the energy distribution of the pair, $\delta_k$ is the shift due to the Coulomb interaction, and $\varepsilon_k$ is the quasiparticle spectrum. The parameters $u_k$ and $v_k$ of (21) satisfy the conditions $u_k^2 = (1/2)\left(1 + \xi_k/\sqrt{\xi_k^2 + \Delta_k^2}\right)$, $v_k^2 = (1/2)\left(1 - \xi_k/\sqrt{\xi_k^2 + \Delta_k^2}\right)$. The expressions (23)-(27) are the result of a simple substitution of the Bogolubov transformation into the Hamiltonian (20). Similar expressions have been previously derived in the numerous papers (see, for instance, Ref. 1,2, 5-23). The dynamic Stark effect and the consequent increase of the order parameter were studied by Comte and Mahler[7], while the effect of a strong resonance field was discussed in Ref. 18 and 19.

The order parameter can be found as a solution of the equation

$$\delta_k u_k v_k - (u_k^2 - v_k^2)\Delta_k = 0 , \qquad (28)$$

Its numerical solutions for a wide range of parameters were studied by many authors[1,2, 5-23]. Typical values of the order parameter are in the range of a few meV. The experimentally estimated value was $\Delta \sim$ (1-2) meV [24-27]. However, the experiments were carried out at room temperature $T \approx 25$ meV $\gg \Delta$.

In the BCS theory of the superconductivity the critical temperature is of the order of the energy gap $\Delta$. This is because Cooper pairs are formed by electrons, which locate near the Fermi surface, where many free spaces exist. An energy about $\Delta$ is enough for the destruction of any Cooper pair. A similar situation exists in the case of *e-h* condensation in a semiconductor[1,2,5-23], when the carrier density is not large and the photogeneration of pairs occurs far inside the bands near the Fermi energy (see Eq. (13)).

The features of the system under study are a) the interaction of *e-h* pairs with the electromagnetic field happens within a narrow spectral band at the band gap; b) a very large carrier density and strong degeneracy of the semiconductor at room temperature. Indeed, the Fermi energy of electrons and holes are $\varepsilon_F^e = (3\pi^2 N)^{2/3}/m_e \approx 350$ meV and $\varepsilon_F^h \approx 45$ meV at the carrier density $\sim 6 \times 10^{18} cm^{-3}$. As a result, if a pair gets an additional energy of about $\Delta \sim$ 2-4 meV, it cannot break up into the electron and hole, since all states, which are available for transitions, are occupied by unpaired electrons and holes [27, 28] (see Fig. 1).

For the description of this situation, it is required to determine an approximate wave function of the condensed state. In order to do this, we use an analogy with the wave function of the vacuum state of Bogolubov quasiparticles

$$|\text{vac}> = \prod (u_k - v_k a_k^+ b_{-k}^+)|0> . \qquad (29)$$

Here $|0>$ is the state of intrinsic vacuum of the semiconductor, i.e. the absence of electrons and holes. The wave function (29) can be used as a first approximation of the ground state of the *e-h* condensate in a semiconductor. It describes low-density plasma, when the number of excitations is small [2, 5-7, 33].

In our case the condensed state exists in combination with a broad layer of a normal Fermi plasma of electrons and holes having larger energies. Unpaired particles block all possible channels of decay of bound *e-h* pairs of the condensate [27]. It is this reason that is responsible of a dramatically enhanced critical density of the condensed state. This fact must be taken into account in the wave function of the ground state.

Because approximate eigenstates of the system look like $\alpha_{k_i}^+ \alpha_{k_j}^+ \beta_{k_n}^+ |\text{vac}>$, it is worth to choose the wave function of the ground state as

$$|\phi_0> = \prod_{k,k'} \alpha_k^+ \beta_{k'}^+ |\text{vac}> , \qquad (30)$$

where the multiplication is evaluated for all occupied electron $k$ and hole $k'$ states, which are determined by the energy distribution of the injected carriers with corresponding Fermi energies.

The destruction of *e-h* pairs happens due to collisions with optical phonons with simultaneous transitions of electrons and holes to states with larger energies. The Hamiltonian of the interaction can be written as

$$H_{ph} = V_Q^e a_{k+Q}^+ a_k + V_Q^h b_{k+Q}^+ b_k ,$$

(31)

where $V_Q^{e,h}$ are the operators acting on the wave function of the phonon, $Q$ is its corresponding wave number. Let consider now a probability of a transition of an electron with the wave number $k$ from the initial state (the ground state of the system $|\phi_0>$) to the final state

$$|\phi_f>=\alpha^+_{k+Q}\alpha_k|\phi_0>$$

(32)

with the wave number $k+Q$. The amplitude of the transition is

$$A_{k,k+Q}=<V_Q^e>u_{k+Q}u_k<\text{vac}|\alpha_k\alpha_k^+\alpha_{k+Q}\alpha_{k+Q}^+\alpha_k\alpha_k^+|\text{vac}>, \qquad (33)$$

where $<V_Q^e>$ is the matrix element taken for the phonon wave function. The squared matrix element should be multiplied by the occupation numbers of the initial and free final states $v_k^2(1-f_{k+Q}^e)$.

Thus, the probability of the transition of an electron from the state $k$, which is paired with a hole $-k$, to the state $k+Q$ with the absorption of a phonon is

$$W^e(k,k+Q)=\frac{2\pi}{\hbar}(N_Q+1)|M_{k,Q}|^2 u_k^2 v_k^2 u_{k+Q}^2(1-f_{k+Q}^e)\times \\ \delta(E^e(k+Q)-E^e(k)-\hbar\Omega) \qquad (34)$$

where $N_Q=(\exp[\hbar\Omega/T]-1)^{-1}$ is the phonon distribution number. The energy of the optical phonon in GaAs $\hbar\Omega \approx 36\,\text{meV}$ can be considered to be practically constant for all phonons having wave numbers about Fermi one $k_F \sim 10^7 cm^{-1}$. $f_{k+Q}^e$ is the Fermi function of the standard distribution of unpaired electron in the state $k+Q$. For estimations, we take the value $|M_{k,Q}|^2$ to be equal to the value of Fröhlich Hamiltonian

$$|M_Q|^2=\frac{2\pi e^2}{VQ^2}\hbar\Omega\left(\frac{1}{\varepsilon_\infty}-\frac{1}{\varepsilon_0}\right) \qquad (35)$$

where $V$ is the volume of the crystal, $\varepsilon_0, \varepsilon_\infty$ are the static and high-frequency dielectric constants.

For the determination of the total probability of the scattering of the electron from the given state $k$ one should evaluate the sum (34) for all possible final states

$$W_k^e = \sum_{k'} W^e(k,k') . \qquad (36)$$

Since the sum in (36) evaluates for states of unpaired electrons with the energies $\varepsilon_{k'}^e \gg \Delta$, the energy of the final state is $\varepsilon_{k'=k+Q}^e \approx \hbar\Omega$ and the value of the given wave number is $Q \approx k' = (2m_e\Omega/\hbar)^{1/2}$.

Finally we have ($u_{k+Q} \approx 1$)

$$W_k^e = 2 \times \frac{1}{\sqrt{2}} \frac{r_0(m_0c^2)^{3/2}(\hbar\Omega)^{1/2}}{\hbar^2 c} \left(\frac{1}{\varepsilon_\infty} - \frac{1}{\varepsilon_0}\right)\left(\frac{m_e}{m_0}\right)^{1/2} u_k^2 v_k^2 \times$$
$$\left(\exp(\hbar\Omega/T) - 1\right)^{-1} \times \left[1 - \left(\exp\left[(\hbar\Omega - \mu_e)/T\right] + 1\right)^{-1}\right] \qquad (37)$$

Similarly, for the total probability of the hole scattering from the state $k$ we have

$$W_k^h = 2 \times \frac{1}{\sqrt{2}} \frac{r_0(m_0c^2)^{3/2}(\hbar\Omega)^{1/2}}{\hbar^2 c} \left(\frac{1}{\varepsilon_\infty} - \frac{1}{\varepsilon_0}\right)\left(\frac{m_h}{m_0}\right)^{1/2} u_k^2 v_k^2 \times$$
$$\left(\exp(\hbar\Omega/T) - 1\right)^{-1} \times \left[1 - \left(\exp\left[(\hbar\Omega - \mu_h)/T\right] + 1\right)^{-1}\right] \qquad (38)$$

In (37) and (38) the number 2 accounts for the spin degeneracy of the electron (hole).

It is obvious from the expressions (37) and (38) that the destruction of the *e-h* condensate occurs mainly due to scattering of heavy holes on optical phonons. Indeed, $\mu_e \gg \mu_h$ and the factor of existence of a free space $(1 - f_{k+Q}^{e,h})$ ensures larger values of $W_k^h$ compared with $W_k^e$.

Figure 4 illustrates the dependence of $W(T)$. The values of (37) and (38) are relatively small at low enough temperatures. The condensate is stable for some time. Due to its resonant interaction with propagating electromagnetic field, the latter establishes coherency and condensation occurs faster and faster. The collective coherent *e-h* state finally decays radiatively in a superradiant manner, and we observe the emission of large femtosecond pulses [24-28]. As the temperature increases, the value of (38) begins to be larger than the critical value. Yet the cooperative state has no enough time to be formed, the collective decay does not happen, and finally *e-h* ensemble recombines spontaneously. In this case one see a low-intensity nanosecond output pulse.

It seems to be pretty obvious to determine the critical temperature of the condensate $T_c$ by the condition of equality of the characteristic decay time of the *e-h* condensate $W^{-1}(T)$ to the characteristic incubation time of superradiance $\tau_s$

$$W(T_c)\tau_s \sim 1 \tag{39}$$

It is well-known from the theory of superradiance [34,35] that the incubation time is determined by the characteristic Dicke time $\tau_D = \frac{8\pi S T_1}{3\lambda^2 N}$, where $S$ is the emission cross section, $N$ is the number of particles in the sample, $T_1$ is the spontaneous relaxation time, and $\lambda$ is the emission wavelength. An approximate relation between these times gives [34,35]

$$\tau_s \sim \tau_D \left(\frac{1}{2}\ln N\right)^2 \tag{40}$$

Using (40), one can get a generally universal dependence $\tau_s \sim 1/N$. However, when the superradiant pulse width is much shorter that the transit time,[35] the accuracy of the relation (40) is too poor. By the numerical solution of full system of Maxwell-Bloch equations, $\tau_s$ was estimated to be 1-2 ps for our case.[35]

Figure 5 shows dependencies of the critical temperature on the *e-h* density obtained using the condition (39). The characteristic superradiant time $\tau_s$ was taken to be 1, 1.5, and 2 ps for the carrier density $3\times 10^{18} cm^{-3}$. The corresponding time for other densities was recalculated using the relation $\tau_s \sim 1/N$. It is clearly seen that the critical temperature is much larger than the parameter $\Delta \sim 2-4\,\text{meV}$ and can exceed room temperature. The critical temperature $T_c$ increases sharply with the concentration and is relatively small at densities $<2\times 10^{18} cm^{-3}$ that correspond to lasing conditions in GaAs/AlGaAs heterostructures.[36] Therefore, estimated values of the critical temperature are in qualitative agreement with available experimental data [24-28].

V. CONCLUSION

We have presented in this paper the qualitative model, which can explain the results of the experiments [24-28], where the *e-h* condensation was observed at room temperature. We have shown that in case of larger carrier densities and strong degeneracy of the semiconductor the critical temperature of the condensation is determined by the Fermi energy of the plasma components. This is conditioned by the necessity for the electron (hole) of a pair to perform a transition to one of free states, which locate at the Fermi energy.[27] Such transition needs an energy much larger compared with the order parameter $\Delta$.

This should not be considered as a contradiction to previous theories of *e-h* condensation because we deal here with a substantially different situation. The energy region of the quasiparticles interaction between each other and with the resonant electromagnetic field in our case lies far below the Fermi energy.

We have also demonstrated that in dense degenerate *e-h* plasma there exists an energy level for the bound *e-h* state, which locates right below the band gap. This level approaches the boundary of the quasicontinuous spectrum of unpaired particles of the plasma as the density increases. This low-energy level is virtually empty under normal conditions and at room temperature since $\Delta \ll kT$. It is the resonant electromagnetic field, which plays a crucial role in filling the level up and the condensation of composite bosons (bound *e-h* pairs) on to it. The electromagnetic field can be either internal recombination field, or an external laser field with properly chosen wavelength and amplitude.

The critical temperature of the condensation in the system under study is determined by the relation (39), which connects the probability of breaking a bound pair up with the characteristic superradiant incubation time. The estimated values of $T_c$ are in good agreement with available experimental data.

## Acknowledgements

The authors would like to thank H.Kan and T.Hiruma of Hamamatsu Photonics K.K. for the financial support and Yu.M.Popov of PN Lebedev Physical Institute for the helpful discussions.

* Electronic address: peter@mail1.lebedev.ru

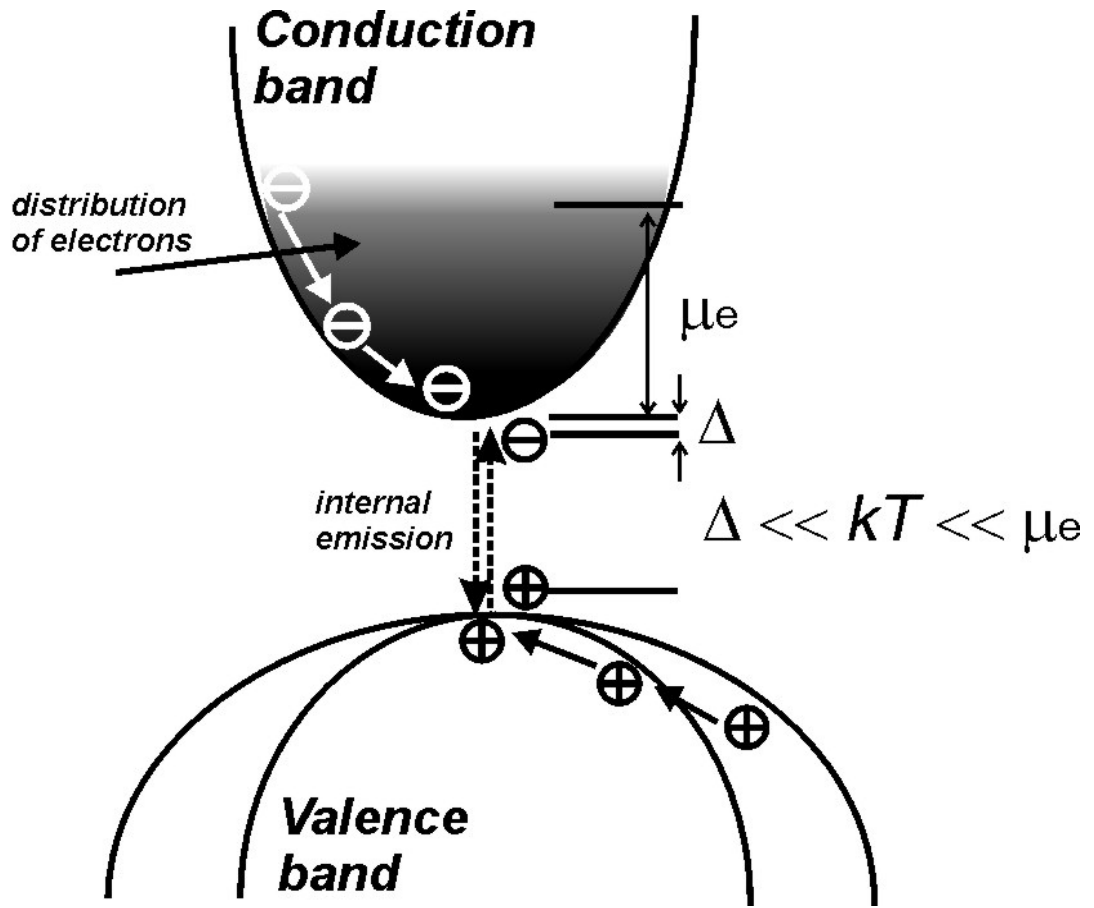

Figure 1. Band structure diagram of a semiconductor with degenerate large-density *e-h* plasma.

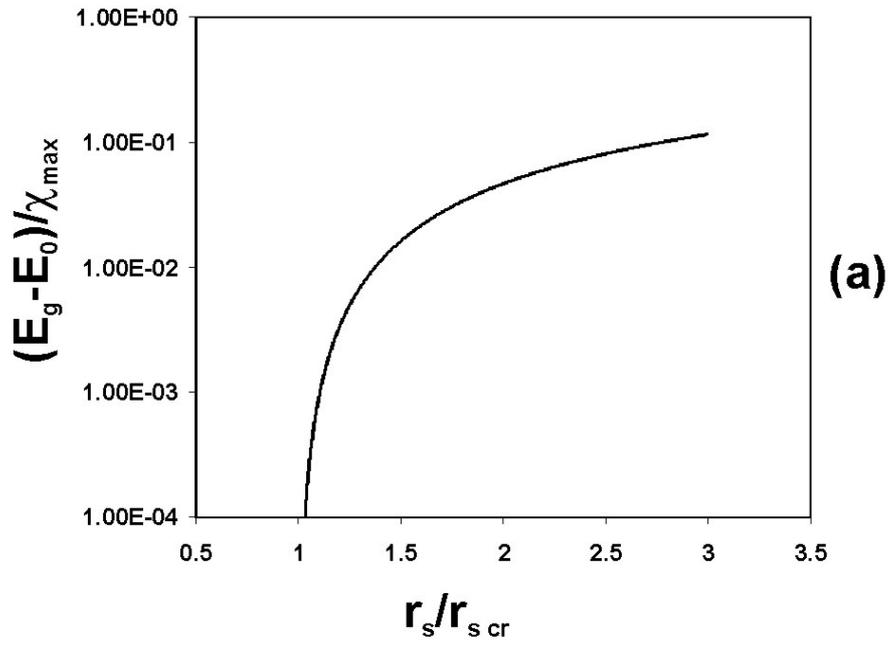

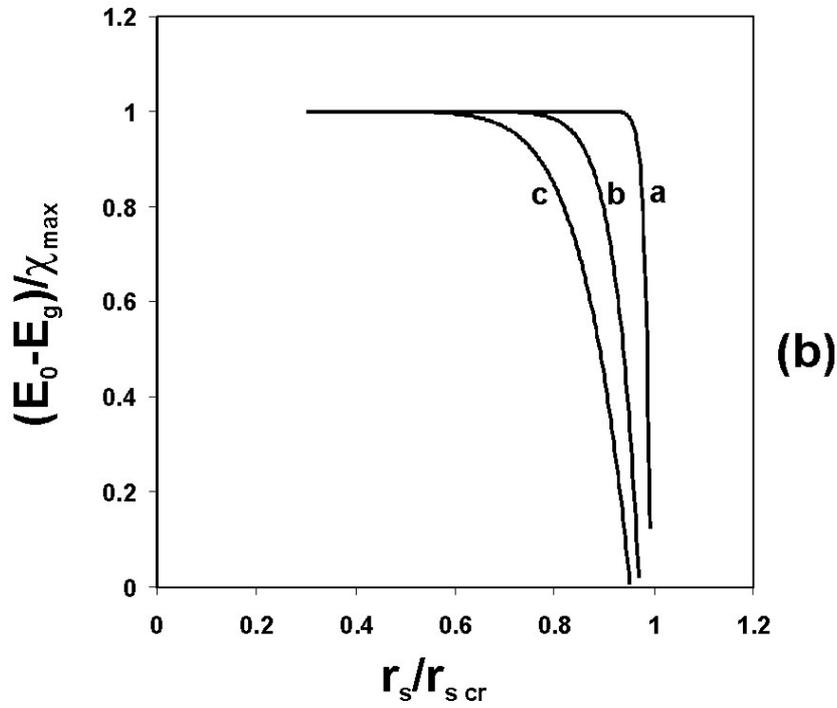

Figure 2. Position of the energy level of the bound *e-h* state against the mean interparticle distance $r_s$. a) the level lies below $E_g$ (for an infinitely large crystal), the solution of Eq. (7); b) the level lies between $E_g$ and the actual bottom of the band, determined by the minimum dimension of the crystal, a) $\delta = 10^{-4}$, b) $\delta = 2.5 \times 10^{-3}$, c) $\delta = 10^{-2}$.

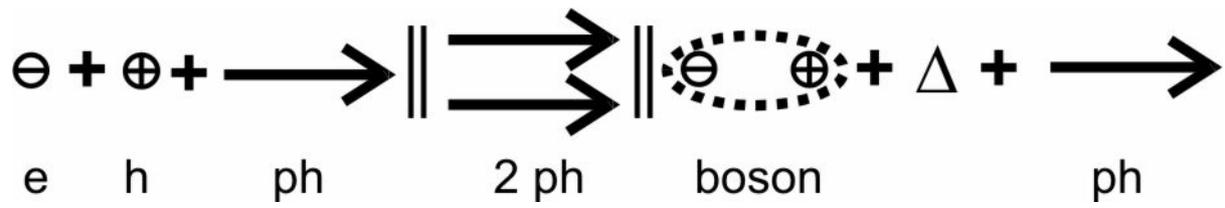

Figure 3. Proposed mechanism of the condensation with the mediated action of photons. *e* is an electron, *h* is a hole, *ph* is a photon, Δ is the bound energy.

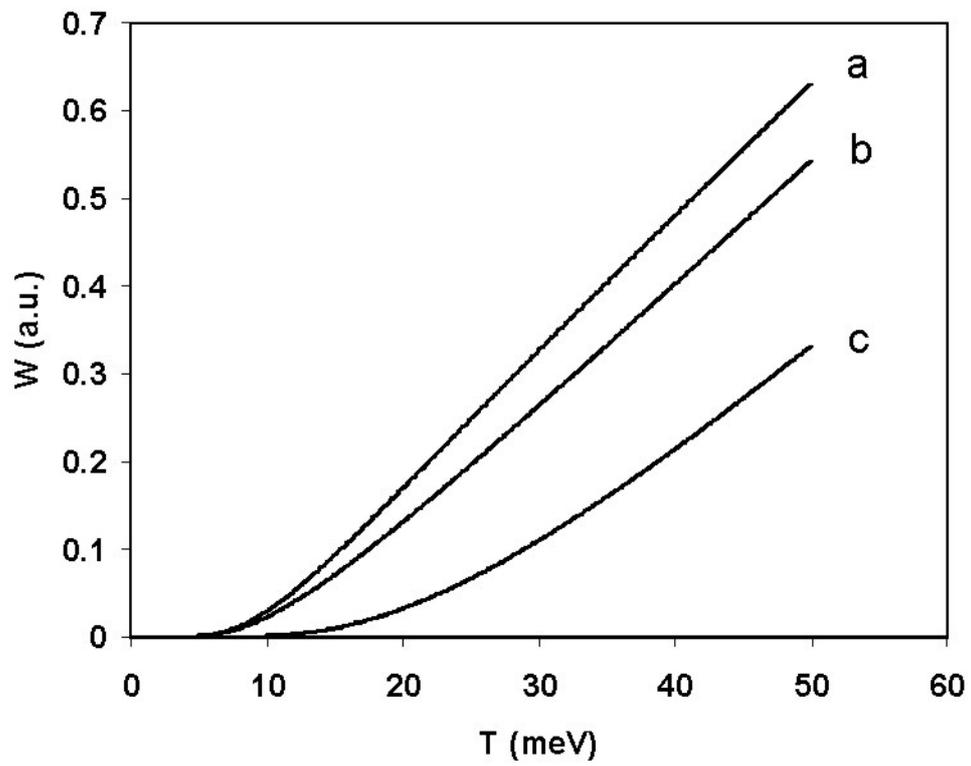

Figure 4. The dependence of the probability of the destruction of *e-h* pairs on temperature at different values of $r_s$, a) $r_s = 1$, b) $r_s = 0.5$, c) $r_s = 0.3$.

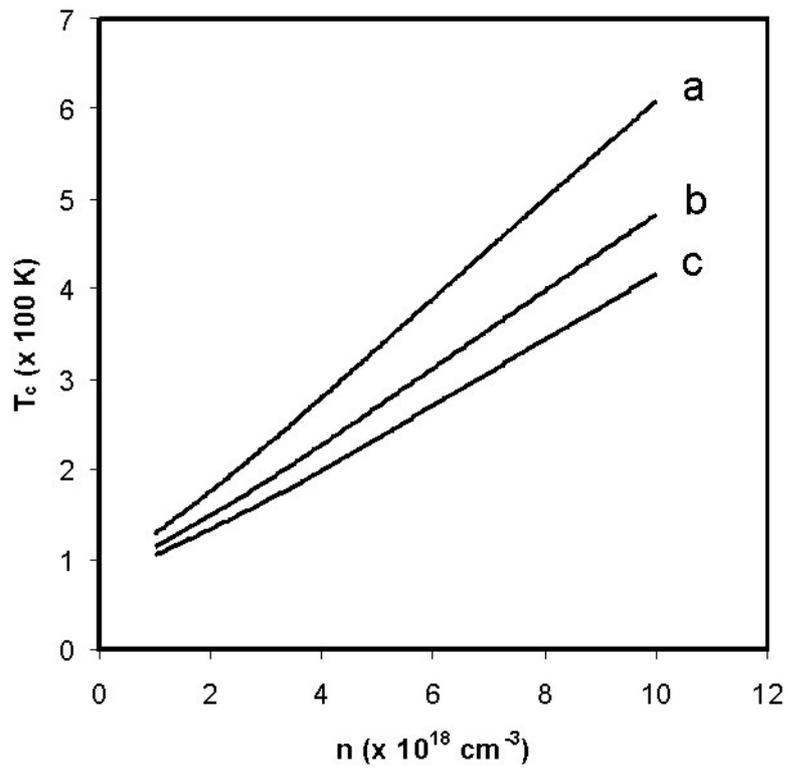

Figure 5. The critical temperature against the *e-h* density for different values of $\tau_s$ a) $\tau_s = 1$ ps, b) 1.5 ps; c) 2 ps